\documentclass[prb,superscriptaddress,aps]{revtex4}
\usepackage{graphicx}
\usepackage{dcolumn}
\usepackage{bm}

\usepackage{color}

\begin{document}


\bigskip\bigskip\bigskip

\title{Self consistent thermal wave model description of the transverse dynamics
for relativistic charged particle beams in magnetoactive plasmas}

\author{Renato Fedele}
\email{renato.fedele@na.infn.it} \affiliation{Dipartimento di
Scienze Fisiche, Universit\`{a} Federico II and INFN Sezione di
Napoli, Complesso Universitario di M.S. Angelo, via Cintia,
I-80126 Napoli, Italy}

\author{Fatema Tanjia}
\email{tanjia@na.infn.it} \affiliation{Dipartimento di Scienze
Fisiche, Universit\`{a} Federico II and INFN Sezione di Napoli,
Complesso Universitario di M.S. Angelo, via Cintia, I-80126
Napoli, Italy}

\author{Sergio De Nicola}
\email{sergio.denicola@ino.it} \affiliation{Istituto Nazionale di
Ottica - C.N.R., Pozzuoli (NA), Italy} \affiliation{Dipartimento
di Scienze Fisiche, Universit\`{a} Federico II and INFN Sezione di
Napoli, Complesso Universitario di M.S. Angelo, via Cintia,
I-80126 Napoli, Italy}

\author{P. K. Shukla}
\email{ps@tp4.rub.de} \affiliation{Center of Advanced Studies in
Physical Sciences, Ruhr-Universität Bochum, Bochum, Germany}

\author{Du\v san Jovanovi\'c}
\email{djovanov@ipb.ac.rs} \affiliation{Institute of Physics,
University of Belgrade, Belgrade, Serbia}

\centerline {\textsf{P5.006 - 38th EPS Conference on Plasma
Physics, Strasbourg, France, 26 June - 1 July, 2011}}

\bigskip\bigskip

\maketitle

The plasma wake field (PWF) excitation \cite{Pisin-Chen} is one of
the efficient mechanisms to generate, in a plasma, ultra-intense
electric and magnetic fields. To produce these fields, a
relativistic electron/positron beam is launched into the plasma.
Due to the very strong nonlinear and collective beam-plasma
interaction, the beam becomes the driver of a large amplitude
plasma wave that follows the beam with almost its own speed
(\textit{plasma wake}) and carries out both longitudinal and
transverse electric fields (\textit{plasma wake fields}). If the
beam length is much greater than the plasma wavelength
(\textit{long beam limit}), the entire beam experiences the
effects of the wake fields that itself has produced (\textit{self
interaction}).

We study the self interaction of an electron/positron beam
travelling in a magnetoactive, collisionless, cold plasma in the
overdense regime, i.e., \small{$n_0 \gg n_b$}, (\small{$n_0$} and
\small{$n_b$} being the unperturbed plasma and beam densities,
respectively). We assume that: a strong constant and uniform
external magnetic field acts along the z-axis, \small{$B_0 = B_0
\hat{\mathbf{e}}_z$}; the electron/positron beam is travelling
along the z-axis, i.e. \small{$\mathbf{v}_b = \beta c
\hat{\mathbf{e}}_z$} (\small{$\beta \simeq 1$}); the ions are at
rest to form a background of positive charge. Moreover, we
consider a fluid model of the \textit{beam-plasma} system,
characterized by the electron fluid velocity,
\small{$\textbf{u}(\mathbf{r},t)$,} the electron plasma and beam
number densities, \small{$ n(\mathbf{r},t)$} and
\small{$\rho_b(\mathbf{r},t)$}, respectively, the electron plasma
and the beam current densities, \small{$-en\mathbf{u}$} and
\small{$\beta q \rho_b c\, \hat{\mathbf{e}}_z$}, respectively,
where $q=-e$ and $q=e$ for electrons and  positrons, respectively.
We ignore the longitudinal beam dynamics and concentrate our
investigation on the transverse effects only. We express the total
electric and magnetic fields of the beam-plasma system in terms of
the four-potential \small{$(\mathbf{A},\phi)$} and introduce small
deviations of all the quantities with respect to the initial state
\small{$n = n_0,\, \textbf{u}=0$, $\textbf{A}=
\textbf{A}_0(\textbf{r})=\left(\textbf{B}_0\times
\textbf{r}\right)/2$, $\phi = 0$}; then we write
\small{$\textbf{u}= \hat{\mathbf{e}}_z u_{1z} +
\textbf{u}_{1\perp}$, $\textbf{A}= \textbf{A}_0 +
\hat{\mathbf{e}}_z A_{1z}+ \textbf{A}_{1\perp}$, $\phi=\phi_1$}.
Furthermore, we assume the long beam limit, i.e.
\small{$\partial^2/\partial\xi^2 \ll \omega^2_{p}/c^2\equiv k_p^2$
($\omega_{p}$} being the electron plasma frequency), which under
suitable boundary conditions implies that
\small{$|\mathbf{A}_{1\perp}|\ll A_{1z}$} and
\small{$|\mathbf{u}_{1\perp}|\ll u_{1z}$}. Consequently, after
linearizing the Lorentz-Maxwell system of equations, we obtain the
equation for the dimensionless \textit{wake potential}\, $U_w$,
driven by the beam density, i.e., \small{
$\left(\nabla^2_\perp-\kappa^2\right) U_w=
\kappa^2\rho_b/n_0\gamma_0 $}, where \small{$\omega_{UH}$} is the
upper hybrid frequency, \small{$\kappa\equiv
\omega_{p}^2/\omega_{UH}c^2$}, \small{$\nabla_\perp$} is the
transverse part of the gradient operator, and {\small$  U_w =
\left(A_{1z}-\phi_1\right)/ m_0\gamma_0c^2 $} with \small{$A_{1z}=
A_{1z}(\textbf{r}_\perp,\xi)$} and \small{$\phi_1 =\phi_1
(\textbf{r}_\perp,\xi)$} the perturbations of both the
longitudinal component of the vector potential and the electric
potential, respectively. Here, \small{$\textbf{r}_\perp$} is the
transverse position vector, \small{$\xi = z-\beta c t \simeq
z-ct$} plays the role of time-like variable, \small{$m_0$} and
\small{$\gamma_0$} are the rest mass and the unperturbed
relativistic gamma factor of the single particle in the
electron/positron beam, respectively. To provide a self consistent
description of the transverse electron/positron beam dynamics, we
have to consider the total force acting on the single particle of
the beam, that accounts for the transverse gradient of $U_w$ as
well as the effects produced by the external magnetic field
$\mathbf{B}_0$. To this end, we describe the transverse beam
dynamics by means of the Thermal Wave Model
(TWM)\,\cite{TWM1,TWM2}.

TWM provides an effective description of the charged-particle beam
dynamics in terms of a complex function, say $\Psi$, called beam
wave function (BWF). The transverse spatio-temporal evolution of
the BWF is given by {\small$ i\epsilon\partial\Psi/\partial\xi =
\mathcal{H}\left(\mathbf{r}_\perp,-i\epsilon\nabla_\perp,\xi
\right)\Psi $}, where, for normalized \small{$\Psi$}, {\small
$\rho_b(\textbf{r}_\perp,\xi) =
(N/\sigma_z)|\Psi(\textbf{r}_\perp,\xi)|^2$} (\small{$N$} and
\small{$\sigma_z$} being the total number of particles and the
beam length, respectively) \cite{TWM1,TWM2},
\small{$\mathcal{H}(\textbf{r}_\perp, \textbf{p}_\perp, \xi)$} is
the effective Hamiltonian describing the perturbed transverse
motion of a single particle of the beam, and \small{$\epsilon$}
the transverse beam emittance. Provided that the longitudinal
dynamics is ignored, i.e. \small{$\textbf{p}= \hat{\mathbf{e}}_z
m_0\gamma_0 c + \textbf{p}_{1\perp}$}, it is easily seen that,
\small{$\mathcal{H}=\Delta H/H_0=(H-H_0)/H_0$} is the relative
first-order perturbation of the single electron/positron
Hamiltonian \small{$H(\textbf{r}, \textbf{p},
\xi)=c\sqrt{(\mathbf{p}-\frac{q}{c}\mathbf{A})^2+m^2_0c^2}+q\phi$},
where \small{$H_0=m_0\gamma_0c^2$} is the initial unperturbed
total energy of the single particle of the beam. Consequently, the
above equations for BWF and \small{$U_w$} can be cast as the
following \textit{Zakharov system} of equations, viz.,
{\small\begin{eqnarray} i\epsilon\frac{\partial
\psi_m}{\partial\xi}=
-\frac{\epsilon^2}{2}\frac{1}{r_\perp}\frac{\partial}{\partial
r_\perp}\left(r_\perp\frac{\partial \psi_m}{\partial
r_\perp}\right)+U_w\psi_m
+\left(\frac{1}{2}Kr^2_\perp+\frac{m^2\epsilon^2}{2r^2_\perp}\right)\psi_m.
\label{b7}\
\end{eqnarray}
\begin{eqnarray}
\frac{1}{r_\perp}\frac{\partial}{\partial
r_\perp}\left(r_\perp\frac{\partial U_w}{\partial
r_\perp}\right)-\kappa^2 U_w= \kappa^2
\frac{N}{n_0\gamma_0\sigma_z}\,|\psi_m|^2, \label{z1-ter}\
\end{eqnarray}}
where we have put \small{$\Psi (r_\perp,\varphi,\xi)=
\exp\left[im\left(\varphi-k_c\xi/2\right)\right]\psi_m(r_\perp,\xi)$}
with \small{$m$} integer and
\small{$K=(\omega_c/2\gamma_0c)^2\equiv\left(qB_0/2m_0\gamma_0c^2\right)^2\equiv
\left(k_c/2\right)^2$}. This system governs the self consistent
\textit{spatio temporal} evolution of the \textit{PWF
self-interaction} of the electron/positron beam. In principle,
once eq.(\ref{z1-ter}) is solved for $U_w$, we get the functional
$U=U\left[\,|\psi_m|^2\,\right]$ that makes eq. (\ref{b7}) the
generalized nonlinear Schr\"{o}dinger equation (NLSE). Note that,
due to the \textit{helicity} phase factor
\small{$\exp\left[im\left(\varphi- k_c\xi/2\right)\right]$} for
\small{$m\neq 0$} the BWF describes vortices states associated
with the orbital angular momentum of the beam particles in the
external magnetic field (\small{$m$} plays the role of
\textit{vortex charge}). Based on the pair of eqs. (\ref{b7}) and
(\ref{z1-ter}), an investigation, both analytical and numerical,
has been carried out, taking into account the diverse limiting
cases. Hereafter, and in Ref. \cite{P5.021}, we give a summary of
the maim results.

\begin{figure}[htb]
\centering
\includegraphics[width=110mm]{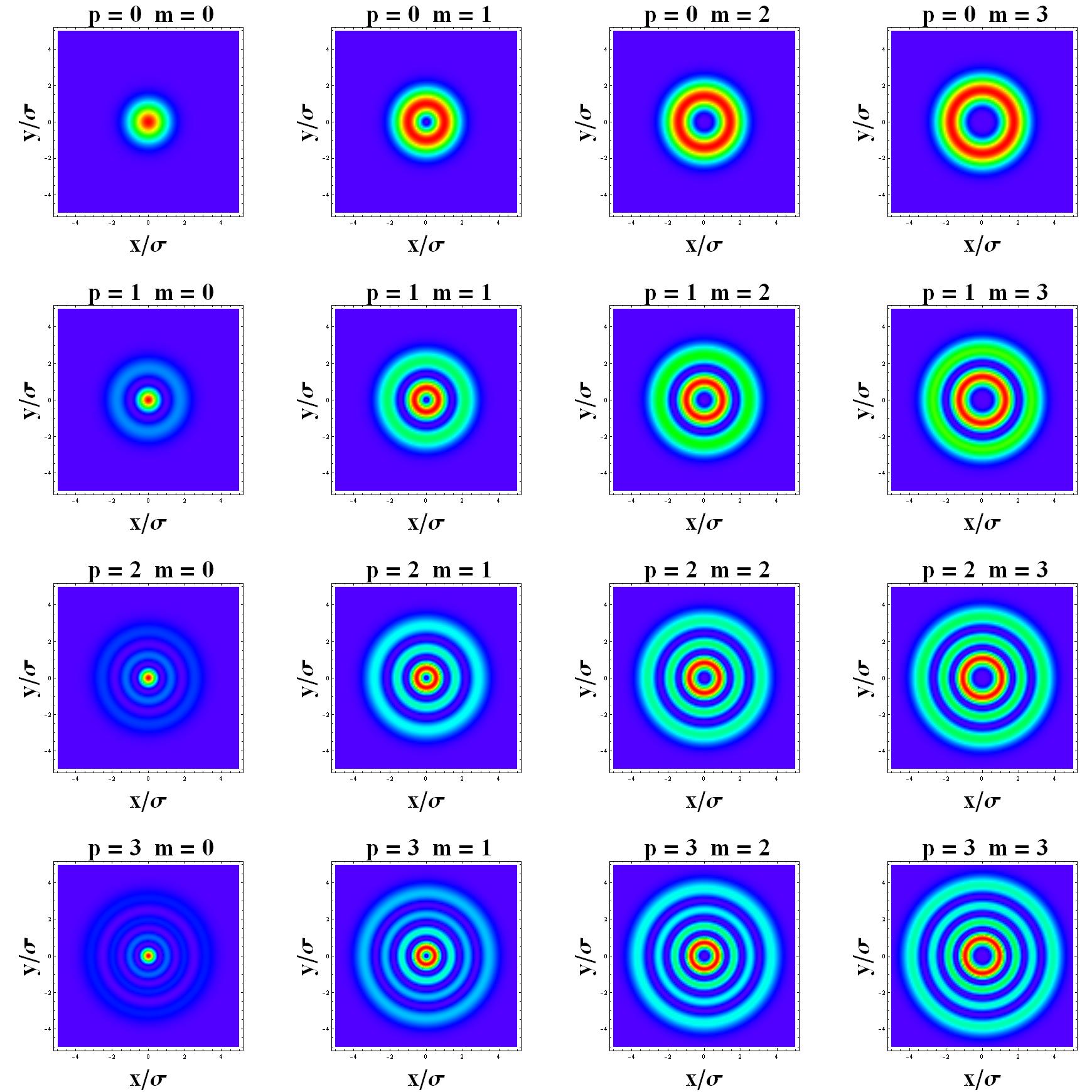}
\,\,\,
\includegraphics[width=45mm]{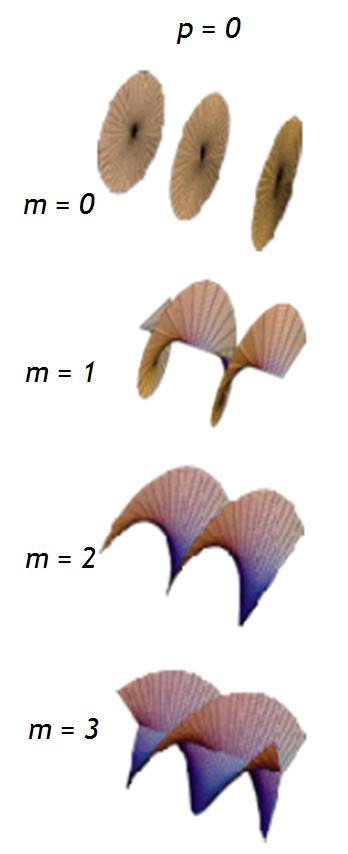}
\caption{\small\textit{Left: Density plots for $|\Psi_{m,p}|^2$ in
the normalized plane ($x/\sigma$, $y/\sigma$) for some
combinations of $p$ and $m$. Right: qualitative representation of
the BWF phase for different values of the vortex charge $m$ when
$p=0$.}}\label{Fig1}
\end{figure}

First of all, in the linear limit, i.e.
\small{$U=U\left[\,|\psi_m|^2\,\right]\simeq 0$} (the beam is
travelling along $\mathbf{B}_0$ and the self interaction is
assumed negligible), for arbitrary integer $m$, the pair of eqs.
(\ref{b7}) and (\ref{z1-ter}) reduces to a linear Schr\"{o}dinger
equation that admits solutions in the form of Laguerre-Gauss
modes, say \small{$\psi_{p,m}(r_\perp,\xi)$} (\small{$p$} is an
arbitrary integer), whose transverse size,
\small{$\sigma_{p,m}(\xi)$}, is proportional to the one associated
with the purely Gaussian mode (\small{$m=p=0$}), say
\small{$\sigma(\xi)$}. The latter satisfies the following envelope
equation: \small{$d^2\sigma/d\xi^2 +
K\sigma-\epsilon^2/\sigma^3=0$}. For a long enough region along
$z$, it describes the typical \textit{sausage-like} transverse
beam size modulations (betatron oscillations), preserving the
collapse (focusing to a single point). Fig. \ref{Fig1} shows the
density plot for \small{$\left|\psi_{p,m}\right|^2$} in the
normalized plane (\small{$x/\sigma$, $y/\sigma$}) for some
combinations of $p$ and $m$. For a given $m$, the number of rings
increases as $p$ increases. When $m = 0$, the structure of the
density plots is always constituted by a central core plus p
rings, whilst for $m\neq 0$, the density plots are associated with
the diverse states of vortices.\\

When the self interaction is not negligible, one can suitably
investigate two limiting cases.\\
(i). If the transverse beam size is much greater than the plasma
wavelength, i.e. \small{$k_p\sigma\gg1$}, the system (\ref{b7})
and (\ref{z1-ter}) reduces to the 2D \textit{Gross-Pitaevskii
equation}, i.e., {\small $$i\epsilon\frac{\partial
\psi_m}{\partial\xi}=
-\frac{\epsilon^2}{2}\frac{1}{r_\perp}\frac{\partial}{\partial
r_\perp}\left(r_\perp\frac{\partial \psi_m}{\partial
r_\perp}\right)
-\frac{N}{\sigma_zn_0\gamma_0}\mid\psi_m\mid^2\psi_m
+\left(\frac{1}{2}Kr^2_\perp+\frac{m^2\epsilon^2}{2r^2_\perp}\right)\psi_m\,.
$$} In Ref.\cite{P5.021} we carry out a numerical analysis of this sort of
NLSE, showing the existence of vortices, nonlinear coherent states
(2D solitons) and beam halos. here, from the virial equation
associated to this NLSE, we find the following envelope equation:
\small{$$d^2\sigma_m^2/d\xi^2+4K \sigma_{m}^2
=4\mathcal{A}_m\,,$$} where {\small $\mathcal{A}_m
\equiv\pi\epsilon^2
\int_0^\infty\left[\left|\partial\psi_m/\partial
r_\perp\right|^2+(m^2/r_\perp^2)\left|\psi_m\right|^2\right]r_\perp
dr_\perp - (\pi N/\sigma_z
n_0\gamma_0)\int_0^\infty\mid\psi_m\mid^4r_\perp dr_\perp +
(K/2)\sigma_m^2$} is constant of motion, i.e. {\small
$d\mathcal{A}_m/d\xi = 0$}, and {\small $\sigma_m \equiv
\left[2\pi\int_0^\infty r_\perp^2 |\psi_m
(r_\perp,\xi)|^2r_\perp\,dr_\perp \right]^{1/2}$} is the
transverse beam size. Note that the interplay between the positive
and the negative terms in the definition of {\small
$\mathcal{A}_m$} can make this quantity positive, negative or
zero. If {\small $\mathcal{A}_m>0$,} in correspondence of the
initial conditions {\small $\sigma_0 \equiv\sigma_m(0)$,} and
{\small $\sigma'_0 \equiv\left(d\sigma_m/d\xi\right)_{\xi=0}$,}
the envelope equation describes stable oscillations in {\small
$\sigma_m$} with frequency {\small $2\sqrt{K}$,} in the range
{\small
$0<\frac{1}{2}\sigma_0'\,^2+\frac{1}{2}K\sigma_0^2<\mathcal{A}_m
$,} whilst {\small $\sigma_m$} would reach zero in a finite time
in the range {\small $0<\mathcal{A}_m
<\frac{1}{2}\sigma_0'\,^2+\frac{1}{2}K\sigma_0^2 $.} If {\small
$\mathcal{A}_m <0$,} always {\small $\sigma_m$} would reach zero
in a finite time. The latter cases would correspond to a collapse
instability. However, as {\small $\sigma_m$} reaches very small
values, the condition {\small $k_p\sigma_m\gg1$} is no longer
satisfied and the collapse does not take place.
It is interesting to describe the self-interaction of an
electron/positron beam that enters a thin plasma slab
(\textit{plasma lens}) of length $l$ at $\xi=0$ where, as initial
conditions, is assumed that \small{$\sigma'_0=0$} and the BWF is
purely Gaussian,
\small{$\psi(r_\perp,0)=\exp[-r_\perp^2/2\sigma_0^2]/\sqrt{\pi}\sigma_0$}.
Provided that$\sqrt{K}l\ll 1$, the self-interaction is very short,
namely the the BWF remains almost unchanged except for the
appearance of a \textit{chirping phase factor}, that accounts for
a strong change in the particle momentum distributions in the
transverse plane (\textit{kick approximation}),
viz.,\small{$\psi(r_\perp,l)\simeq\psi(r_\perp,0)
\exp[ir_\perp^2/2\epsilon\rho(l)+i\phi(l)]$}, where
\small{$\rho(\xi)$} and \small{$\phi(\xi)$} are the curvature
radius of the wavefront and a homogeneous phase at location $\xi$,
respectively, such that, as \small{$\xi\rightarrow 0$},
\small{$\rho\rightarrow \infty$} and \small{$\phi\rightarrow 0$}.
At the lens exit (\small{$\xi=l$}), the transverse beam size is
\small{$\sigma(l)\simeq \sqrt{\sigma_0^2 + 2\left(\mathcal{A}_0 -
K\sigma_0^2\right)l^2}$}\,(\small{$\mathcal{A}_0$} being the
constant of motion corresponding to the Gaussian beam).
Consequently, under the condition $\mathcal{A}_0-K\sigma_0^2<0$,
the beam is focussed and out of the lens it reaches a minimum spot
size that is greater than zero, according to the envelope equation
in the vacuum (i.e., no plasma and $K=0$), viz.,
\small{$d^2\sigma/d\xi^2 -\epsilon^2/\sigma^3 =0 $}. We have
\textit{weak focusing} (\textit{strong focusing}) if the above
condition is satisfied for positive (negative) values of
\small{$\mathcal{A}_0$}.\\
(\textbf{ii}). If \small{$k_p\sigma\ll 1$}, eqs. (\ref{b7}) and
(\ref{z1-ter}) reduces to a nonlocal NLS equation whose
aberrationless (i.e. Gaussian) approximate solution leads to the
envelope equation \small{$d^2\sigma/d\xi^2 + K\sigma +\eta/\sigma
-\epsilon^2/\sigma^3 =0 $}, where \small{$\eta =
\left(2e^2\omega_p^2N/m_0\gamma_0\omega_{UH}^2\sigma_z\right)$}.
It admits the \textit{self-equilibrium solution}
\small{$\sigma'_{eq}= \sqrt{-\mu+\sqrt{1+\mu^2}}\,\,\sigma_{eq}$},
where \small{$\sigma_{eq}\equiv \epsilon^{1/2}/K^{1/4}$} is the
corresponding self-equilibrium solution for the linear case
\small{$\eta=0$} and \small{$\mu =\eta/2\epsilon\sqrt{K} $}. It is
easily seen that \small{$\sigma'_{eq}<\sigma_{eq}$} in all the
range \small{$0\leq\mu<\infty$}. This implies that the term
proportional to \small{$1/\sigma$} accounts for the squeezing of
the beam.

\end{document}